\documentclass[aps,twocolumn,showpacs,showkeys,superscriptaddress,amssymb]{revtex4}

\date{\today}
\usepackage{graphicx,epsfig,latexsym,overpic,amssymb,color}
\usepackage{booktabs,longtable}

 \preprint{\today}

\begin{document}
\title{ Helium-cluster decay widths of molecular states in beryllium and carbon isotopes}
\author{J.C. Pei}  \affiliation{School of Physics and MOE Laboratory of Heavy Ion Physics, Peking
University, Beijing 100871, China}
\author{F.R. Xu}
\email{frxu@pku.edu.cn} \affiliation{School of Physics and MOE
Laboratory of Heavy Ion Physics, Peking University, Beijing
100871, China} \affiliation{Institute of Theoretical Physics,
Chinese Academy of Sciences, Beijing 100080, China}
\affiliation{Center of Theoretical Nuclear Physics, National
Laboratory of Heavy Ion Collisions, Lanzhou 730000, China}

\date{\today}

\begin{abstract}
The $\alpha$ particle and $^6$He emissions from possible
molecular states in beryllium and carbon isotopes have been
studied using a mean-field-type cluster potential. Calculations
can reproduce well the $\alpha$-decay widths of excited states in
$^{8}$Be, $^{12}$C and $^{20}$Ne. For the nucleus $^{10}$Be, we
discussed the $\alpha$-decay widths with different shapes or decay
modes, in order to understand the very different decay widths of
two excited states. The widths of $^{6}$He decay from $^{12}$Be
and $\alpha$ decays from $^{13,14}$C are predicted, which could be
useful for future experiments.

\end{abstract}


\maketitle

The experimental search for molecular-type structures in light
nuclei is of increasing interest. The Ikeda diagram as a guide
line reveals that cluster structures are expected to appear near
decay thresholds~\cite{ikeda68}. Evidences for dimers in beryllium
isotopes and polymers in carbon isotopes have been summarized by
von Oertzen~\cite{oertzen}. It is convincing that the candidate
cluster bands should have very large moments of inertia,
corresponding to molecular-like shapes. Cluster states above
thresholds should be understandable though it is difficult to
perform the precise measurements of cluster decay widths. The
model calculations of decay widths are generally dependent on the
deformations and angular momenta of nuclear states. Hence, the
predictions of widths can provide useful structure information
about tentative cluster states.

In $^{8}$Be, the rotational cluster band and its decay property
have been well established based on an $\alpha$-$\alpha$
structure~\cite{buck77}. In heavier beryllium isotopes, two
$\alpha$ particles and additional valence neutrons can give rise
to covalent molecular binding. Recent experiments have observed
the $\alpha$+$^6$He decay in $^{10}$Be~\cite{liendo,freer06}and
the $^6$He+$^6$He decay in $^{12}$Be~\cite{saito04,freer99}, which
would be the indication for molecular states. Different molecular-type
structures in $^{10}$Be and $^{12}$Be have been suggested by the
antisymmetrized molecular dynamics (AMD)
calculations~\cite{kanada-en'yo03,kanada-en'yo99}. It is also
intriguing that carbon isotopes with three $\alpha$ clusters can
have two different configurations: triangular and linear. For
$^{13}$C and $^{14}$C, cluster bands based on the
$\alpha$+$^{9}$Be and $\alpha$+$^{10}$Be systems have been
proposed~\cite{oertzen,milin02,oertzen04}, respectively. The
experimental study of $\alpha$ decays from excited states of
$^{13,14}$C is attracting interest~\cite{price06, soic04}.

Many theoretical works based on the WKB approach have shown the
successful calculations of decay life-times,
e.g.~\cite{buck90,delion,renzz,mohr,ohkubo,Ch06,Du06}. Combining
the Bohr-Sommerfeld quantization, Buck {\it et al.} has achieved
the calculations of spectra of cluster states using macroscopic
cluster potentials~\cite{buck04}. Recently, we suggested a
mean-field-type cluster potential for various charged-cluster
decays from the ground states (g.s.) of even-even heavy
nuclei~\cite{jcpei}. In the present work, we extended our
calculations by including the deformations and angular momenta of
cluster states. We focus the decay properties of excited states in
light nuclei, particularly in beryllium and carbon isotopes, which
are attracting great interest in experiments.

The cluster potential in the quantum tunnelling approach can be
written as (e.g.,~\cite{buck90}),
\begin{equation}
V(r)=V_N(r)+V_C(r)+\frac{\hbar ^2}{2\mu r^2}(L+1/2)^2,
\end{equation}
that contains the nuclear potential $V_N(r)$, the Coulomb
potential $V_C(r)$ and the Langer modified centrifugal
potential~\cite{buck90}. We suggested the nuclear potential
$V_N(r)$ by~\cite{jcpei}
\begin{equation}
V_N(r)=\lambda[Z_cv_p(r)+N_cv_n(r)],
\end{equation}
where $\lambda$ is the folding factor; $N_c$ and $Z_c$ are the
neutron and proton numbers of the cluster, respectively; $v_n(r)$
and $v_p(r)$ are the single-neutron and -proton
potentials(excluding the Coulomb potential) respectively, obtained
from Skyrme-Hartree-Fock calculations with the SLy4
force~\cite{chabanat}. The Coulomb potential $V_{\rm C}(r)$ is
well defined physically and should not be folded. We have
approximated the Coulomb potential by $V_{\rm C}(r)=Z_{\rm
c}v_{\rm c}(r)$, where $v_{\rm c}(r)$ is the single-proton Coulomb
potential given by the mean-field calculation. In the spherical
case, the folding factor $\lambda$ is determined with the
Bohr-Sommerfeld quantization condition~\cite{buck90},
\begin{equation}
\int _{r_1}^{r_2} \sqrt{\frac{2\mu}{\hbar
^2}|Q_L^{*}-V(r)|}=(2N+1)\frac{\pi}{2}=(G-L+1)\frac{\pi}{2},
\end{equation}
where $r_1$, $r_2$ (and $r_3$ in Eq.(5)) are the classical turning
points obtained by $V(r)=Q_{L}^{*}$ (the decay energy). The global
quantum number G is estimated by the Wildermuth rule, depending on
the configurations of valance nucleons~\cite{wildermuth}. The
cluster decay energy from an excited state is given by,
\begin{equation}
Q_L^{*}=Q_{0}+E_J^{*},
\end{equation}
where $Q_0$ is the decay energy from the ground state, and
$E_J^{*}$ is the excitation energy of a given state with the spin
$J$. The decay process can occur only if the state has a positive
$Q^{*}_L$ value.

\begin{table}
\caption{The calculated $\alpha$-decay widths of excited states in
$^8$Be and $^{20}$Ne,  with comparison with the experimental
data~\cite{firestone}. }
\begin{ruledtabular}
\begin{tabular}{rcccc}
~~~$J^{\pi}$&$E_J^*$& ($G, L$)&$\Gamma_{\alpha}$(calc.) &
$\Gamma_{\alpha}$(expt.)\\
~~~&[MeV]&&[keV]&[keV]\\
\hline
\multicolumn{5}{c}{$^8$Be ($Q_0$=91.8 keV)}\\
0$^{+}$&0.00~ &(4, 0)& 7.8 ev & 6.8$\pm$1.7 ev\\
2$^{+}$&3.04~ &(4, 2)& 1.6$\times10^3$ & 1.5$\times10^3$\\
4$^{+}$&11.40 &(4, 4)& 2.3$\times10^3$ & $\approx$3.5$\times10^3$\\
\hline
\multicolumn{5}{c}{$^{20}$Ne ($Q_0$=$-$4.729 MeV)}\\
 6$^{+}$&8.78~&(8, 6) & 0.17 & 0.11$\pm$0.02\\
 8$^{+}$&11.95 &(8, 8)& 90 ev &35$\pm$10 ev\\
 1$^{-}$&5.79~~&(9, 1)&18 ev&28$\pm$3 ev\\
 3$^{-}$&7.16~~&(9, 3)&4.7&8.2$\pm$0.3\\
 5$^{-}$&10.26&(9, 5)&54&145$\pm$40\\
 %{\{ 7$^{-}$}&13.69&(9, 7)&29&310$\pm$30\\
 7$^{-}$&15.37&(9, 7)&120&110$\pm$10\\
%{\{ 9$^{-}$}&17.43&(9, 9)&3.2&220$\pm$25\\
9$^{-}$&22.87&(9, 9)&116&225$\pm$40\\
\end{tabular}
\end{ruledtabular}
\end{table}

The partial decay widths can be calculated by~\cite{buck90},
\begin{equation}
\Gamma =P\frac{\displaystyle (\hbar
^2/4\mu)exp[-2\int_{r_2}^{r_3}dr k(r) ]}{\displaystyle \int
_{r_1}^{r_2} dr/2k(r)},
\end{equation}
where $k(r)=\sqrt{\frac{2\mu}{\hbar^2}|Q_L^{*}-V(r)|}$ is the wave
number, and P is the preformation factor of the cluster. For
even-even nuclei, the extreme P=1 assumption under using the
Bohr-Sommerfeld condition can well reproduce the experimental
half-lives of various cluster decays~\cite{buck90,jcpei}. The
decay half-life can be obtained by $T_{1/2}=\hbar$ln2/$\Gamma$.

In the axially deformed case, the decay width can be approximated
by averaging widths at various directions of the space as
follows~\cite{denisov}
\begin{equation}
\Gamma=\int_0^{\pi/2} \Gamma(\theta)\sin(\theta)d\theta.
\end{equation}
To calculate the width $\Gamma({\theta})$, a deformed cluster
potential has to be employed. In the present work, the deformed
potential $V(r, \theta)$ is constructed with axially deformed
single-particle potentials $v_n(r,z)$, $v_p(r,z)$ and $v_c(r,z)$
that are given by the shape-constrained Skyrme-Hartree-Fock
calculation~\cite{pei06}, with the folding factor determined at
the spherical case. The variables ($r$, $z$) are the cylindrical
coordinates, and $\theta$ is the angle between the symmetry axis
and the radius for the cluster emission.

\begin{table}
\caption{ The calculated $\alpha$-decay widths for the
$K^{\pi}=0_2^{+}$ cluster band in $^{10}$Be ($Q_0$=$-$7.413 MeV)
at the different cases of the spherical and deformed shapes.}
\begin{ruledtabular}
\begin{tabular}{rccccc}
 ~~$J^{\pi}$&$E_J^*$[MeV]~~& $\Gamma_{\alpha}$(sph.) & $\Gamma_{\alpha}$(def.)&
 $\Gamma_{\alpha}(z$-axis$)$& $\Gamma_{\alpha}$(expt.) \\
\hline
 ~~0$^{+}_2$&6.18&       &       &        &$-$\\
 ~~2$^{+}$&7.54  &0.51 ev&0.85 ev& 4.1 ev &22$\pm$7 ev~\cite{liendo}\\
 ~~4$^{+}$&10.15 &35 keV &97 keV &584 keV &130 keV~\cite{freer06}\\

\end{tabular}
\end{ruledtabular}
\end{table}

The most well-established cluster structures in light nuclei are
in $^8$Be and $^{20}$Ne~\cite{wildermuth},  with an
$\alpha$-particle coupling to the magic cores of $^4$He and
$^{16}$O, respectively. Their cluster structures can be well
described by the semiclassical cluster model~\cite{buck77,buck95}.
For $^8$Be, we take G=4 for the ground-state band, according to
the Wildermuth condition~\cite{wildermuth}. For $^{20}$Ne, we take
G=8 for the $K^{\pi}$=$0^{+}$ band (the ground-state band) and G=9
for the $K^{\pi}$=$0^{-}$ band, as discussed in
Refs.~\cite{wildermuth, buck95}. The calculated results are given
in Table I. It can be seen that the present calculations agree
well with the experimental widths~\cite{firestone} within a factor
of 3.  $^{20}$Ne is a particularly interesting nucleus that can
have very different structures for different  states. While the
$K^{\pi}=0^{-}$ band with the sequence of $J^{\pi}=1^{-}, 3^{-},
..., 9^{-}$ has an almost pure $\alpha$+$^{16}O$ cluster
structure, the $0^{+}$ ground-state band has a considerable
mixture of the cluster structure and the deformed mean-field
structure~\cite{kimura}. For the $0^{-}$ band, both experiments
and our calculations give remarkable large $\alpha$-decay widths
comparable with the Wigner limit~\cite{kimura}, indicating the
significant $\alpha$+$^{16}O$ cluster structure.

\begin{table}
\caption{ The calculated $^6$He-widths of the states belonging to
the $K^{\pi}=0_3^{+}$ band in $^{12}$Be ($Q_0$=$-10.11$ MeV). The
AMD predictions are also given for comparison. }
\begin{ruledtabular}
\begin{tabular}{cccc}
 ~~$J^{\pi}$&$E_J^*$[MeV]& $\Gamma_{^{6} {\rm He}}$(present)& $\Gamma_{^{6}{\rm He}}$~\cite{kanada-en'yo03}~~\\
  && [keV]&[keV]\\
\hline
 ~~0$^{+}_3$&10.9&410 &700 ~~\\
 ~~2$^{+}$&11.3&285 &1 ~~\\
 ~~4$^{+}$&13.2&190 &7 ~~\\
 ~~6$^{+}$&16.1&34 &16 ~~\\
%    &6$^{+}$(18.6)&320keV&\\
 ~~8$^{+}$&20.9&3.7\\

\end{tabular}
\end{ruledtabular}
\end{table}

For heavier beryllium isotopes, the $2\alpha$-cluster
structures play an important role. For example, the well-known
parity inversion in $^{11}$Be is related to a large prolate
deformation of $^{10}$Be due to a developed cluster
structure~\cite{kanada-en'yo99}. The deformed mean-field
calculation for $^{11}$Be with the mixture of a $sd$ neutron orbit
is insufficient to gain enough binding for the $J^{\pi}=1/2^{+}$
state~\cite{pei06}. Deformations in beryllium isotopes are
dependent on the distance of the two $\alpha$ particles. In
$^{10}$Be, the cluster band with $0_{2}^{+}$(6.18 MeV),
$2^{+}$(7.54 MeV) and $4^{+}$ (10.15 MeV) members have been
established experimentally~\cite{freer06,mmilin05}. It was found
that a channel radius as large as 8$\pm$1 fm has to be adopted to
reproduce the $\alpha$-decay width for the 7.54 MeV
state~\cite{liendo}, while the  $\alpha$-width of the 10.15 MeV
state is associated with a smaller channel radius of 5$\sim$6
fm~\cite{freer06}.

From the shell-model viewpoint, the $^{10}$Be($0_2^{+}$) structure
has a ($sd$)$^2$ configuration due to the large prolate
deformation~\cite{itagaki00}. Hence we assume a global number G=6
for the decay calculation. The calculated widths in $^{10}$Be are
displayed in Table II. It can be seen that calculations in
the spherical case give much smaller $\alpha$-decay widths compared
with experiments. As predicted by the molecular orbital
model~\cite{itagaki00}, the $\alpha$-$\alpha$ distance in the
second $0^{+}$ state is about 4 fm. This corresponds to an axis
ratio of 2.5$:$1 (or a prolate deformation of
$\beta_2\approx1.1$). With such a large deformation employed in
Eq.(6), the $\alpha$-width for the 10.15 MeV state can be
reproduced reasonably. We found that the $\alpha$-width of the
$2^{+}$(7.54 MeV) state is sensitive to the decay energy rather
than the deformation of the cluster potential (the width can be
reproduced with an increase of only 50 keV in the decay energy).
This can be understood considering that, for a state near the
threshold, the cluster tunnelling with a very small decay energy
is dominated by the long tail of Coulomb potential (the tail is
not sensitive to the deformation).

Considering the $\alpha$:$2$n:$\alpha$ molecular structure of the
$0_2^{+}$ band~\cite{freer06}, we estimated the $\alpha$-width by
assuming that the $\alpha$ particle emits along the $z$ axis with
the $\beta_2=1.1$ deformed potential. The calculated widths are
given in Table II. It can be seen that the $\alpha$-width for the
7.54 MeV state is improved significantly. However, the
$\alpha$-width for the $4^{+}$ (10.15 MeV) state is overestimated.
It was pointed out that the $0_2^{+}$ band states could contain
the significant mixture of the $^5$He+$^5$He
configuration~\cite{itagaki00,ito}. The mixture is particularly
remarkable for the 10.15 MeV state because it is almost on the
$^5$He+$^5$He decay threshold, which could result in a reduced
$\alpha$-decay width.

For the neutron-rich $^{12}$Be nucleus, some resonance states were
recently observed, decaying to $\alpha$+$^8$He and $^6$He+$^6$He
channels~\cite{freer99,saito04}. A rotational band with a
$^6$He+$^6$He structure has been suggested to have a very large
momentum of inertia ($\hbar^2/2\Im=0.14$ MeV)~\cite{freer99}. The
AMD calculations interpreted that resonances happen in molecular
states built on the $0_3^{+}$ configuration~\cite{kanada-en'yo03}.
Considering that the highest state of this band would be the
$J^{\pi}=8^{+}$ member~\cite{freer99,saito04}, we take G=8 in the
$^6$He+$^6$He decay calculation. A spherical cluster potential is
assumed. In Table III,  the obtained widths are compared with the
AMD predications~\cite{kanada-en'yo03}. As stated
in~\cite{kanada-en'yo03}, the $^6$He-width of the $0_3^{+}$ state
is very large due to the lack of centrifugal barrier. It is shown
that the present widths significantly decrease with increasing
angular momenta. In these states, $\alpha$+$^8$He decays have also
been observed~\cite{freer99} but their description is beyond the
cluster tunnelling model because of too large decay energies.

\begin{figure}[t]
\vspace{-4pt} \hspace{0pt}
\begin{minipage}[b]{0.5 \textwidth}
\includegraphics[width=7cm,height=11.0cm]{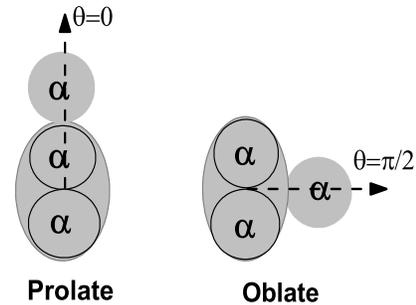}
\end{minipage}
\vspace{-200pt} \caption{\label{fig:separation} The schematic
picture for the two modes of $\alpha$ decays in carbon isotopes,
corresponding to the prolate(linear) and oblate(triangle)
structures, respectively. }
\end{figure}

Carbon isotopes heavier than $^{12}$C can have three-centre
cluster structures with different geometric shapes. As suggested
by von Oertzen~\cite{milin02,oertzen04}, there are possible
prolate(linear) and oblate(triangular) cluster states in carbon
isotopes. The $K^{\pi}=0_{1}^{+}$ and $3^{-}_1$ bands in $^{12}$C
are considered to have triangular configurations. The $0_2^{+}$
state of $^{12}$C has been predicted to be an analogue of Bose
condensate~\cite{tohsaki}. Recently, $\alpha$ decays from the
excited states of $^{13}$C and $^{14}$C have attracted intense
experimental studies (see e.g., \cite{price06, soic04}).

The oblate and prolate configurations certainly lead to different
calculated decay widths. As shown in Fig.1, the cluster potential
radius in the oblate case($\theta=\pi/2$) is smaller than in the
prolate case($\theta=0$). Therefore, decays in oblate case would
be suppressed. In $^{12}$C, the $4^{+}$(14.08 MeV) and
$3^{-}$(9.64 MeV) states belong to the $K^{\pi}=0_{1}^{+}$ and
$3^{-}_1$ oblate bands, respectively. The decay channels of these
low-lying states are nearly pure $\alpha$
decays($\Gamma_{\alpha}\simeq\Gamma$)~\cite{firestone,fedorov},
providing a good chance to study the deformation effect. In Table
IV, with a deformed cluster potential of $\beta_2=0.8$
(corresponding to an axis ratio of 2:1), the obtained
$\alpha$-widths of the $3^{-}$ and $4^{+}$ states at
$\theta=\pi/2$ direction (the oblate case) agree well with the
experimental widths. However, their $\alpha$-widths at $\theta=0$
direction (the linear case) are calculated to be 243 keV for the
$3^{-}$ state and 2.2 MeV for the $4^{+}$ state, which are about
one order of magnitude larger than the oblate calculations. This
indicates that $\alpha$ decays are strongly hindered in triangular
configurations compared with linear configurations.

The $\alpha$ decay from the $0_2^{+}$ state in $^{12}$C has been
studied with the triple cluster model~\cite{fedorov}. This state
has a decay energy of 288 keV to $^{8}$Be(g.s.)+$\alpha$ and an
energy of 380 keV to 3$\alpha$ particles. For the decay to
$^{8}$Be+$\alpha$, the present calculation with a linear structure
($\beta_2=0.8$) gives an $\alpha$-width of 5.9 ev that agree with
the observed value of 8.5 ev. The calculations assuming a
triangular configuration leads to a small width of 1.6 ev. For the
decay to the 3$\alpha$ particles with a triangular structure, our
model estimates a width of 36 ev that is close to the predictions
of the AMD~\cite{kanada-en'yo06} and the triple cluster
model~\cite{fedorov}. However, the contribution of the direct
3$\alpha$-decay process to the total width is less than 4$\%$
obtained in the experiment~\cite{freer94}. Hence, the linear
cluster decay to $^{8}$Be+$\alpha$ should dominate the decay of
the  $0_2^{+}$ state in $^{12}$C.

\begin{table}[t]
\caption{ The calculated $\alpha$-widths of excited states in
carbon isotopes. The experimental total decay width $\Gamma$ are taken
from~\cite{firestone} for $^{12,13}$C and from~\cite{oertzen04} for $^{14}$C.}
\begin{ruledtabular}
\begin{tabular}{lcccc}
 ~~$J^{\pi}$&$E_J^*$[MeV] &($G, L$)& $\Gamma_{\alpha}$(calc.) &  $\Gamma_{}$(expt.) \\
 &&&[keV] &[keV]\\
 \hline
 \multicolumn{5}{c}{$^{12}$C ($Q_0$=$-$7.366 MeV)}\\
 ~~$0_2^{+}$&7.654&(6, 0)&5.9 ev&8.5 ev\\
 ~~3$_1^{-}$&9.641&(5, 3)&17 &34 \\
 ~~4$_1^{+}$&14.08&(4, 4)&158 &258 \\
\hline
 \multicolumn{5}{c}{$^{13}$C ($Q_0=-$10.647 MeV)}\\
 \multicolumn{2}{l}{$K^{\pi}=3/2^{-}$}&&\\
% ~~$3/2^{-}$&9.80&(6, 0)&$-$&26 \\
 ~~$5/2^{-}$&10.82&(6, 2)&0.2 ev&24 \\
 ~~$7/2^{-}$&12.44&(6, 2)&84 &140 \\
 ~~$9/2^{-}$&14.13&(6, 4)&32 &150 \\
 ~~$11/2^{-}$&16.08&(6, 4)&316 &150 \\
%     & $1/2^{+}$(10.996) &(7,1)& 5.7ev&37 keV\\
 \multicolumn{2}{l}{$K^{\pi}=3/2^{+}$}&&\\
 ~~$3/2^{+}$&11.08&(7, 1)&41 ev&$\leq$4 \\
 ~~$5/2^{+}$&11.95&(7, 1)&100 &500 \\
 ~~$7/2^{+}$&13.41&(7, 3)&112 &35 \\
 ~~$9/2^{+}$&15.28&(7, 3)&888 &\\
 ~~$11/2^{+}$&16.95&(7, 5)&125 &330 \\
\hline
 \multicolumn{5}{c}{$^{14}$C ($Q_0=-$12.011 MeV)}\\
 ~~$5^{-}$&14.87&(7, 5)&0.4 &35 \\
 ~~$6^{+}$&16.43&(6, 6)&0.12 &35 \\
% ~~$6^{+}$\tablenotemark[1]      &16.43&(6, 4)& 5 ev &35 keV\\
 ~~$3^{-}$ &12.58&(7, 3)&22 ev&95 \\
 ~~$5^{-}$&15.18&(7, 5)&31 &50 \\
 ~~$7^{-}$&18.03&(7, 7)&14 &70 \\
 ~~$6^{+}$ &14.67&(6, 6)&0.21 &40 \\
\end{tabular}
\end{ruledtabular}
%\tablenotetext[1]{The decay channel of
%$^{14}$C$^{*}$$\rightarrow$$^{10}$Be$^{}(2^{+})$+$\alpha$}
\end{table}

\begin{table}[t]
\caption{ The calculated $\alpha$-widths of the $^{14}$C excited
states decaying to $^{10}$Be(g.s.)+$\alpha$ and
$^{10}$Be$(2^{+})$+$\alpha$ channels.}
\begin{ruledtabular}
\begin{tabular}{lcccc}
  &\multicolumn{2}{c}{18.5 MeV}&\multicolumn{2}{c}{19.8 MeV}\\
~~$J^{\pi}$&$^{10}$Be(g.s.)&$^{10}$Be(2$^{+}$)&$^{10}$Be(g.s.)&$^{10}$Be(2$^{+}$)\\
\hline
~~$4^{+}$& 1.2 MeV & 0.8 MeV & 2.7 MeV&\\
~~$5^{-}$& 0.28 MeV &0.39 MeV &0.7 MeV &1.2 MeV\\
~~$6^{+}$& 11 keV & 30 keV & 41 keV & 230 keV\\
~~$7^{-}$& 0.8 keV & 2.3 keV & 3.5 keV & 25 keV\\
\end{tabular}
\end{ruledtabular}
\end{table}

For $^{13}$C, the $K^{\pi}=3/2^{\pm}$ bands constructed on the
$\alpha$+$^9$Be($3/2^{-}$, g.s.) structure were suggested
experimentally~\cite{milin02}. The parity doublet bands are
related to the reflection asymmetric chain configurations,
corresponding to a very large prolate deformation
($\hbar^2/2\Im=0.19$ MeV)~\cite{milin02}. We assume that the core
(i.e., $^{9}$Be) has a similar deformation to that of
$^{10}$Be with $\beta_2\thickapprox0.6$
experimentally~\cite{iwasaki}. In the $\alpha$-width calculations,
we take G=6 and 7 for the $K^{\pi}=3/2^{-}$ and $3/2^{+}$ bands,
respectively, considering the different parity of the bands. The
angular momenta $L$ that the $\alpha$ particle carries are
obtained by the angular momentum selection among the parent,
daughter and $\alpha$ particle. Furthermore, for the odd nucleus $^{13}$C, 
we take the $\alpha$ preformation factor P=0.6 that was adopted in
the systematical $\alpha$-decay calculations of odd
nuclei~\cite{buck90}. The calculated $\alpha$-widths of the
$K^{\pi}=3/2^{\pm}$ band states in $^{13}$C are listed in Table
IV. The experimental $\alpha$-widths have not been available, but
we give the experimental total widths
$\Gamma$~\cite{milin02,firestone} of the states for comparison.

The calculated $\alpha$-widths for $^{13}$C are in general smaller
than experimental total widths as expected. However, calculated values for
the $11/2^{-}$(16.08 MeV) and $7/2^{+}$(13.41 MeV) states \cite{milin02} are
larger than experimental total widths. In Ref.~\cite{firestone},
the two states were assigned with ($7/2^{+}$) and
($9/2^{-}$), respectively. The recent experiment \cite{price06} 
suggests that the 13.41 MeV state is connected to an oblate
structure and the 16.08 MeV state has a
positive parity. Calculations can be
changed significantly due to the different assignments of spins.

The predicted $\alpha$-widths for $^{14}$C decaying to the ground
state of $^{10}$Be are also presented in Table IV. The bands built
on the $0_2^{+}$(6.589 MeV) and $3_2^{-}$(9.801 MeV) states were
suggested to have oblate (triangular) structures, while bands
built on the $0_3^{+}$(9.746 MeV) and $1^{-}$(11.395 MeV) states
have prolate (chain) structures~\cite{oertzen04}. In Table IV, the
5$^{-}$(14.87 MeV) and 6$^{+}$(16.43 MeV) states have oblate
structures and other states are prolate~\cite{oertzen04}. The
calculated width of the $3^{-}$(12.58 MeV) state is remarkably
smaller than the experimental total width. This would indicate
that the neutron emission channel dominates the decay of the state
that is near the $\alpha$-decay threshold. The decay channel of
$^{14}$C(16.43 MeV)$\rightarrow$$^{10}$Be$^{}(2^{+}, 3.37
$MeV$)$+$\alpha$ has also been observed though it is very
weak~\cite{milin04}. The decay width is estimated to be 5 ev,
giving a small branching ratio compared with the
$^{10}$Be(g.s.)+$\alpha$ decay channel.

It is interesting that, in $^{14}$C, the 18.5 MeV and 19.8 MeV
states have strong channels decaying to both the ground state and
the first $2^{+}$ state of  $^{10}$Be~\cite{milin04,soic04}. No
experimental assignments of spins and parties have been available
for the two states of $^{14}$C. We estimated their $\alpha$-widths
with assuming $J^{\pi}=5^{-}, 7^{-}$(G=7) and
$J^{\pi}=4^{+},6^{+}$(G=6). The calculations are preformed in the
simple spherical shape. As shown in Table V, the calculated
$\alpha$-widths are sensitive to the assignments of angular
momenta. With the spin of 4$\sim$5, broad widths are obtained for
both channels as observed in experiments~\cite{milin04,soic04}.

In summary, we have calculated decay widths for the interesting
cluster states in beryllium and carbon isotopes, such as the
$\alpha$+$^6$He, $^6$He+$^6$He and $^{8,9,10}$Be+$\alpha$ systems,
using a uniform folding cluster potential that has good
descriptions for various cluster radioactivities in heavy nuclei.
It has been shown that the model can well reproduce the
$\alpha$-widths of $^8$Be, $^{12}$C and $^{20}$Ne. The discrepant
decay properties of the 7.54 MeV and 10.15 MeV states in $^{10}$Be
have been discussed. The $^6$He-widths of possible molecular
states in $^{12}$Be and $\alpha$-widths of excited states in
$^{13,14}$C are predicted.

We are grateful to Dr. M. Milin for valuable comments and Dr. P.D.
Stevenson for providing the Skyrme-Hartree-Fock code. This work
has been supported by the Natural Science Foundation of China
under Grant Nos. 10525520 and 10475002, and the Key Grant Project
(Grant No. 305001) of Education Ministry of China. We also thank
the PKU Computer Center where numerical calculations have been
done.

\end{document}